# PyNanospacing: TEM image processing tool for strain analysis and visualization

Mehmet Ali Sarsıl[a], Mubashir Mansoor[b, d], Mert Saraçoğlu[b, c], Servet Timur[b], Mustafa Ürgen[b], Onur Ergen[a*]

[a] Istanbul Technical University, Department of Electronics and Communications Engineering, Istanbul, Turkey

[b] Istanbul Technical University, Department of Metallurgical and Materials Engineering, Istanbul, Turkey

[c] Nanosilver Co. Teknopark İstanbul, Pendik, Istanbul, Turkey

[d] Istanbul Technical University, Department of Applied Physics, Istanbul, Turkey

*Corresponding author: Onur Ergen (oergen@itu.edu.tr)

**Abstract**

The diverse spectrum of material characteristics, including band gap, mechanical moduli, color, phonon and electronic density of states, along with catalytic and surface properties, are intricately intertwined with the atomic structure and the corresponding interatomic bond lengths. This interconnection extends to the manifestation of interplanar spacings within a crystalline lattice. Analysis of these interplanar spacings and the comprehension of any deviations—whether it be lattice compression or expansion, commonly referred to as strain, hold paramount significance in unraveling various unknowns within the field. Transmission Electron Microscopy (TEM) is widely used to capture atomic-scale ordering, facilitating direct investigation of interplanar spacings. However, creating critical contour maps for visualizing and interpreting lattice stresses in TEM images remains a challenging task. Here, we developed a Python code for TEM image processing that can handle a wide range of materials, including nanoparticles, 2D materials, pure crystals, and solid solutions. This algorithm converts local differences in interplanar spacings into contour maps, allowing for a visual representation of lattice expansion and compression. The tool is very generic and can significantly aid in analyzing material properties using TEM images, allowing for a more in-depth exploration of the underlying science behind strain engineering via strain contour maps at the atomic level.

**Program summary**

*Program title:* PyNanospacing
*Developer's repository link:* github.com/malisarsil/PyNanoSpacing
*Licensing provisions:* Creative Commons by 4.0 (CC by 4.0)
*Programming language:* Python 3.11

*Nature of problem:* Strain can significantly alter materials properties and therefore the quantitative characterization of strain at the atomic level is of utmost interest. Although TEM images can provide atomic resolution, a quantitative analysis of lattice strain is a cumbersome task, and this leads to confusions regarding the cause and effect reasoning behind changes in macroscopic materials properties, mainly because atomic scale strains cannot be characterized in a robust manner.

*Solution method:* The PyNanospacing python code is a post-processing library written entirely in Python, which takes as input the TEM images of a material, and calculates the local lattice expansions and compressions. This information is then visualized through strain-contour diagrams.

## 1. Introduction

All material characteristics are essentially dependent on local electron densities (bonds), and because changes in local electron density induce major changes in the material wavefunction, every variable that influence the electron density necessarily alter material properties. The lattice strain, defined as the change in bond lengths among the atoms that make up a crystal lattice, is an important component that determines the electron density in a material. Strain can appear as compression or tension, which results in a decrease or increase in unit cell volume, respectively. Strain can be quantitatively expressed as the relative change in the interplanar spacing ($\varepsilon$), calculated by the Eq. 1.

$$\varepsilon = \frac{(d_p - d_s)}{d_s} \tag{1}$$

Here, $dp$ represents the strained interplanar spacing, and $ds$ is the unstrained interplanar spacing of the bulk (the reference). Strain in a crystal lattice can be caused by mechanical forces (physically induced strain), the presence of vacancies, interstitial/substitutional atoms in solid solutions (chemically induced strain), and structural irregularities such as grain boundaries, quantum confinement, and size effects [1, 2]. Deviation of lattice parameters and interplanar spacing from the bulk counterpart is inevitable in nearly all materials and can be observed across multiple length scales, ranging from macroscopic fractures in metallic alloys and ceramics to thin film interfaces, 2D materials, nanomaterials, and so on. Traditional methods for evaluating such strain in materials involve measuring changes in mechanical or physical qualities rather than seeing changes in interplanar spacing at the atomic scale, mainly due to the cumbersome nature of such analysis.

The variation in the lattice parameter becomes apparent when inspecting a structure using high-resolution transmission electron microscope (HR-TEM) image. However, it is customary to measure the interplanar spacing at a few points only and reporting the average, which fails to capture the comprehensive strain map within the structure, as seen in the prior studies on Au [3], Ag [4], $VO_2$ [5], $Co_3O_4$ [6] nanoparticles $TiO_2$/Si thin films[7]. This shortcoming in TEM analysis is mainly due to the non-trivial and cumbersome nature of such analysis, and the requirement for robust image processing tools that do not yet exist.

The aim of this study is to develop and present a robust solution for creating strain maps based on TEM images. We present PyNanospacing, an open-source software designed to help researchers create accurate strain contour maps from any TEM image. The software is completely written in Python (version 3.11) and examines images of nanomaterials, solid solutions, and heterostructures through the use of several image processing methods to assess the compression and expansion of spacing between ordered planes. All packages are trivial to install. PyNanospacing, which can be installed from the PyPi code repository with pip3 install pynanospacing, is expected to be a valuable addition to the condensed matter Physics and materials research toolkit for the development of strain engineered materials by visualizing interplanar expansion and compression at the atomic scale.

## 2. Mathematical Formalism Behind Image Processing Algorithm

This section offers in-depth information on the mathematical operations and the associated functions from the libraries utilized for the execution of this code. The primary library employed for implementing the majority of the image processing functions is OpenCV, which stands for the Open Source Computer Vision Library. OpenCV is a software library for computer vision and machine learning, provided as open-source software [8].

### 2.1. Image Magnification

The Image library's "resize" function is a powerful tool that allows users to create resized images with specific dimensions. It accepts parameters such as the target size in pixels, an optional resampling filter, a definition for a source image region, and a reducing gap for optimization. The choice of resampling filters, such as Resampling.NEAREST, Resampling.BOX, Resampling.BILINEAR, Resampling.HAMMING, Resampling.BICUBIC, or Resampling.LANCZOS, is adaptable based on the image's mode and bit depth, ensuring appropriate filtering for image magnification. For instance, "Resampling.NEAREST" filter picks one nearest pixel from the input image while ignoring all other input pixels. A demonstration of how this filter performs the magnification can be seen in **Figure 1.**

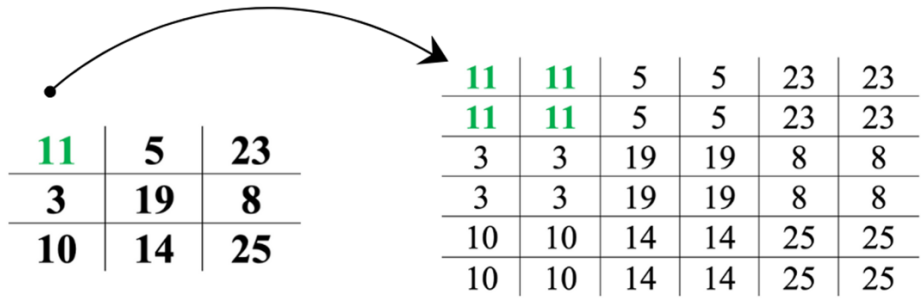

*Figure 1. Demonstration of Resampling.NEAREST Filter for Image Magnification.*

### 2.2. Image Rotation

The rotation of an image is based on fundamental mathematical principles involving linear transformations and matrices. A 2D rotation matrix is employed to rotate an image. The rotation matrix, often denoted as R, is used to transform the coordinates of points in the image, which operates as follows.

Given an initial point (x, y), the transformed point (x', y') after rotation can be calculated using the following equations:

$$x' = x * cos(\theta) - y * sin(\theta) \qquad (2)$$

$$y' = x * sin(\theta) + y * cos(\theta) \qquad (3)$$

In **Eq.2** and **Eq.3** $(x, y)$ represents the original coordinates of a pixel in the image, $(x', y')$ denotes the new coordinates after rotation, and $\theta$ is the rotation angle. The equations involve trigonometric functions, $cosine(cos)$, and $sine(sin)$. The $cosine$ function represents the transformation along the x-axis, while the $sine$ function represents the transformation along the y-axis. The OpenCV library automates this mathematical foundation by calculating the necessary rotation matrix based on the specified rotation angle and image centre. It then applies this matrix to each pixel in the image to perform the rotation, resulting in the rotated image.

### 2.3. RGB to Gray Conversion

The conversion of an RGB (Red, Green, Blue) color image to grayscale is rooted in fundamental principles of image processing and the human visual system [9]. Grayscale images are essentially a single-channel representation of color images, and the transformation is carried out by considering the luminance or brightness of each pixel in the original image. The mathematical foundation for this conversion is typically based on the following formula:

$$Grayscale\ Value(G) = 0.299 * Red\ (R) + 0.587 * Green\ (G) + 0.114 * Blue\ (B) \qquad (4)$$

This formula applies weights to the R, G, and B channels to approximate the perceived brightness of the colors. The coefficients 0.299, 0.587, and 0.114 are derived from the fact that the human eye is more sensitive to green wavelengths and less sensitive to blue [9].

In the context of image processing, the OpenCV library provides a convenient function, cv2.cvtColor, to perform RGB to grayscale conversion. Grayscale images are widely used for various purposes, including feature extraction, edge detection, and simplifying image analysis tasks, thanks to their efficient single-channel representation of visual content [10].

### 2.4. Operations using kernels and strides

Kernels, often represented as small matrices, are employed for various operations on images, such as filtering and feature extraction. They are used in both 1D and 2D forms to process data efficiently. In 1D applications, kernels are applied along a single dimension, while in 2D applications, they work with images, applying operations across both width and height. Strides, on the other hand, dictate the step size at which kernels traverse the data, influencing the output's spatial resolution [11]. **Figure 2** demonstrates a 2D convolution operation wherein a 2 x 2 kernel is moved across the spatial data with a stride of 1 within the provided image.

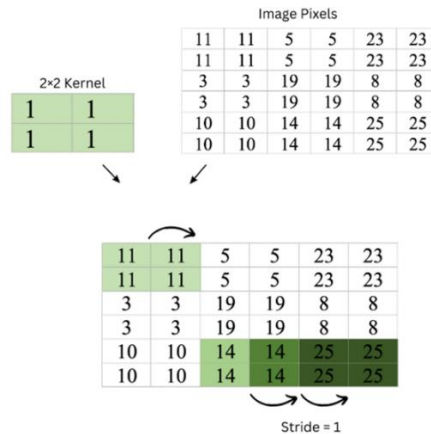

***Figure 2.** Illustration of 2D Convolution Operation with a 2x2 Kernel and Stride of 1 [9].*

## 2.5. Statistical parameter extraction over pixel values: Mean and Skewness

"Average" and "skew" are two essential statistical concepts for understanding and characterizing data distributions.
The "average", often referred to as the "mean," provides insight into the central value of a dataset. It is calculated by summing all the data points and dividing by the total number of points. The mean serves as a measure of typical or central tendency within the data.

$$Mean = \sum \frac{X_i}{n} \quad (5)$$

On the other hand, "skewness" or "skew" describes the asymmetry of a data distribution. It indicates whether the data is skewed to the left (negatively skewed), where the tail extends more to the left, or to the right (positively skewed), where the tail extends more to the right. A perfectly symmetric distribution has zero skew. Skewness is a valuable statistic for understanding the shape and characteristics of data distributions, and can be calculated using the Eq. 6, where $\tilde{\mu}_3$ denotes the skewness value, $N$ is the number of variables in the distribution, $X_i$ is the random variable, $\bar{X}$ is the mean of the distribution and $\sigma$ denotes the standard deviation of the distribution.

$$\tilde{\mu}_3 = \frac{\sum_i^N (X_i - \bar{X})^3}{(N-1) * \sigma^3} \quad (6)$$

## 2.6. Dilation and Erosion

"Dilation" and "erosion" are often applied to binary images, where pixels are either foreground (object) or background (non-object).

Dilation is a process that expands or thickens the foreground regions in a binary image. It involves sliding a small, predefined window called a "structuring element" over the image. When the center of this window aligns with a foreground pixel, the output pixel becomes foreground. Dilation is useful for tasks like joining nearby objects and filling gaps in binary images. It enhances and strengthens the object features within the image.

Erosion, in contrast, is an operation that shrinks or thins the foreground regions in a binary image. It is also applied using a structuring element. When the center of the structuring element aligns with a foreground pixel, the output pixel becomes foreground only if all the pixels within the structuring element are also foreground. Erosion helps remove noise, isolating individual objects, and separating touching objects in binary images. These operations can be used individually or in combination, depending on the specific image processing goals [12]. Cv2.erode and cv2.dilate functions of the OpenCV library are used to allow for the application of erosion and dilation operations.

## 2.7. Skeletonize Function

The "skeletonize" function performs a thinning operation on binary images, reducing the thickness of object regions while preserving their essential features. It iteratively erodes the object regions until they are reduced to their central lines or "skeletons." What sets skeletonization apart is its ability to maintain the topological properties of objects, preserving features like branches, endpoints, and connectivity within them. This makes it ideal for tasks such as shape analysis, pattern recognition, and character recognition, as it simplifies object representation while retaining critical structural information [13]. Various skeletonization algorithms exist, offering different strategies for thinning objects, and Python libraries like OpenCV and sci-kit image provide convenient functions for applying skeletonization to binary images.

### 2.8. Gaussian Smoothing

Gaussian smoothing, often referred to as Gaussian blur, is an objective of reducing noise and refining image quality. This technique is founded on the Gaussian function, which is a bell-shaped curve that acts as a filter to weight pixel values in the smoothing process. The core purpose of Gaussian smoothing is noise reduction, particularly in eliminating high-frequency noise from images, resulting in a cleaner and more visually appealing output. It accomplishes this through convolution, where each pixel in the image is averaged with its neighboring pixels based on the Gaussian kernel's weights. Gaussian kernel in 2D is expressed by the following formula.

$$\frac{1}{2\pi\sigma^2} e^{\left(\frac{-(x^2+y^2)}{2\sigma^2}\right)} \tag{7}$$

The extent of blurring, which determines the trade-off between noise reduction and detail preservation, is regulated by the size of the Gaussian kernel and its standard deviation [14].

### 3. Computation workflow for TEM image analysis

This tool provides a flexible method as it involves several parameters users can input to improve the output image and contour map. The entire procedure followed can be divided into three consecutive steps, which are illustrated in **Fig. 3**.

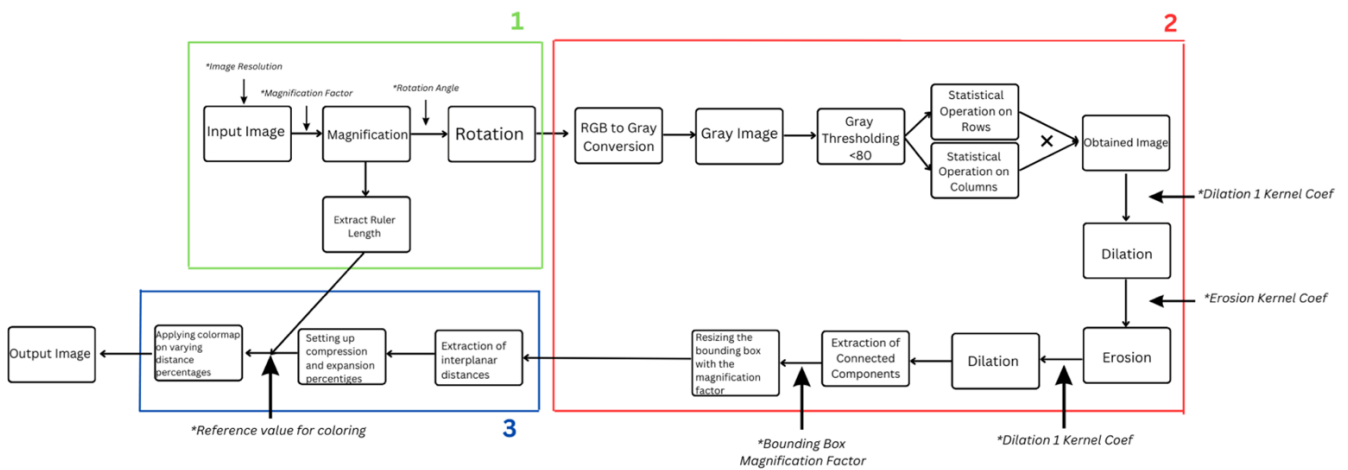

*Figure 3.* *The image processing steps for extracting interplanar spacing and the user-input requirements are described. Parameters denoted with a "*" sign indicate the required user inputs. The entire process can be divided into three primary stages.*

### 3.1. Pre-processing of the image

The user may provide an input image with low resolution, where each unit distance on the image represents a small number of pixels, resulting in minimal pixel count changes for varying distances. This can lead to lower-quality distance-based coloring and limits the application of statistical operations. To address this, the user can specify a parameter that enlarges the image by multiplying its height and width with this factor. Typically, a higher parameter value will improve distance resolution, but it is advisable to find an optimal value to avoid excessive computation time in subsequent algorithmic steps. Additionally, the image contains a horizontal ruler (scale bar) that indicates the visual length of the image's resolution. This ruler is used to determine the pixel count corresponding to its length, as it exhibits a continuous intensity over consecutive pixels, achieved by identifying the longest series of black pixels in the horizontal direction. The user inputs the resolution of the input image in nm, allowing the algorithm to calculate the real-life distance represented by one pixel in nanometers. This value is later used in the color mapping function. Moreover, the input image may not exhibit a vertical alignment of atomic orderings, potentially resulting in imprecise horizontal distance measurements. This is because the horizontal distance between two atom orderings with non-vertical orientations could yield varying incorrect distance values. To address this issue, users can input an angle parameter to transform and rotate the image into a vertical orientation.

### 3.2. Extraction of Region of Interest (ROI)

The enlarged and rotated image is converted to grayscale, resulting in pixel intensities ranging from 0 to 255. Grayscale thresholding is applied by setting a predefined pixel intensity value of 80, which produces a black-and-white image, highlighting the black regions. The resulting image is then fed into another function crucial for extracting the region of interest.

Two additional parameters, namely stride and kernel size, are adjusted by the user to create a one-dimensional kernel that moves along each row of the image. As the kernel traverses the row, it computes the average intensities of pixels falling under it. Furthermore, a skew parameter is calculated in addition to the pixel intensity averages. This process leads to segmenting the pixels into groups based on their average intensity. Segments with an average intensity greater than a predefined threshold of 0.3 and an average skew value less than the overall average of skew values for all segments are retained as white, while all other segments in all rows are turned black. This operation is performed both horizontally and vertically. The resulting image after the row operation is elementwise multiplied by the image after the column operation. These three statistical parameters and their interplay facilitate the identification of the true nanoparticle region while removing unwanted artifacts from the image.

The resulting image may contain many white segments that are vertically and horizontally aligned closely but they may not be connected. To connect these segments, three sequential steps are applied. First, a dilation operation is carried out to expand the white components in the image. Second, an erosion operation is performed to reduce the unwanted components. Lastly, another dilation operation is applied to the image. The overall dilating and eroding impacts of these three steps on the white components depends on the specific image being processed. Users can define different kernel sizes for each operation based on their needs. For example, a noisy image with many small unwanted white components, other than the central region of interest, can be cleaned up by using a smaller erosion kernel, whereas larger noise components can be eliminated by a larger erosion kernel before the initial dilation operation. The final dilation operation primarily aims to enlarge the region of interest. The OpenCV library provides a function for locating connected white components in a binary image, determining their center points, and drawing bounding rectangular boxes around them with the maximum width and height. The bounding box surrounding the connected white regions forming the region of interest is then used in the third step of the overall algorithm.

### 3.3. Interplanar distance extraction and color mapping

In this step, a method is employed to estimate the distance between atomic lines. The bounding boxes obtained are in binary format, where thick white lines represent the atomic orderings. To obtain the most information about the interplanar distance, it is necessary to assume that these lines are thinner. To achieve this, the skeletonize function is used to create a one-pixel-wide representation of the lines (atomic arrays). Subsequently, the image's rows are processed to calculate and store the horizontal black-to-black distances within the bounding box. A custom color mapping function, which determines color variations based on a user-defined reference distance value in nanometers (nm), is applied to color the black-to-black distances. The user provides the reference interplanar spacing value, and the previously calculated parameter for distance per pixel is used to establish its relationship to the user's input. The compression and expansion rates are expressed in percentages relative to the entered reference value. The resulting percentage values are then stored in an Excel file, which includes information about bins and counts of percentages. The colored bounding box is overlaid in its original location on the image, which is rotated back to its original orientation using the inverse of the initial angle. The image is visually smoothed using a Gaussian smoothing filter and then saved to a local folder along with a corresponding color bar.

### 4. Demonstration

A TEM image of a platinum nanoparticle is utilized to illustrate the entire algorithm. The starting image, acquired through Transmission Electron Microscopy (TEM), is presented in the initial image, measuring $128 \times 128$ in size. It was subsequently magnified by a factor of 7. To ensure horizontal distance calculations between vertically aligned atomic orderings, a rotation of 20 degrees was applied on the image. Following this, the image is transformed into a binary format, employing a pixel intensity threshold of 80 to identify and extract black regions within the original image. The outcomes of these operations are showcased in **Fig. 4**.

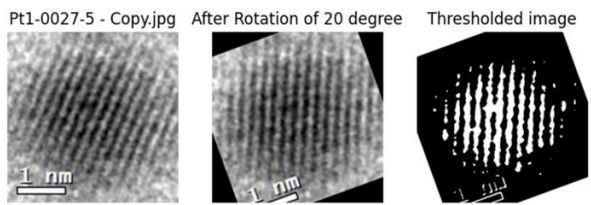

*Figure 4.* Original TEM image of Platinum Nanoparticle, image after rotation, and image after thresholding, respectively. The white regions in the resulting thresholded image aim at extracting the black regions in the original image.

Following the binary thresholding, the resulting image contains regions that are not part of the region of interest (ROI). To eliminate these unwanted regions, three consecutive steps are executed. The same statistical operation is performed both horizontally and vertically, resulting in the second image presented in **Fig. 5**. The kernel size and stride for row operations are determined by two coefficients: the kernel size is calculated by multiplying the image's width by a specified coefficient, while the stride is determined by multiplying the kernel size by another coefficient. These coefficients are set at 0.4 and 0.6, respectively. Subsequently, three operations are applied on this binary image, namely dilation with a kernel size of 21, erosion with a kernel size of 53, and another dilation with a kernel size of 121. The resulting images are depicted in **Fig. 5**.

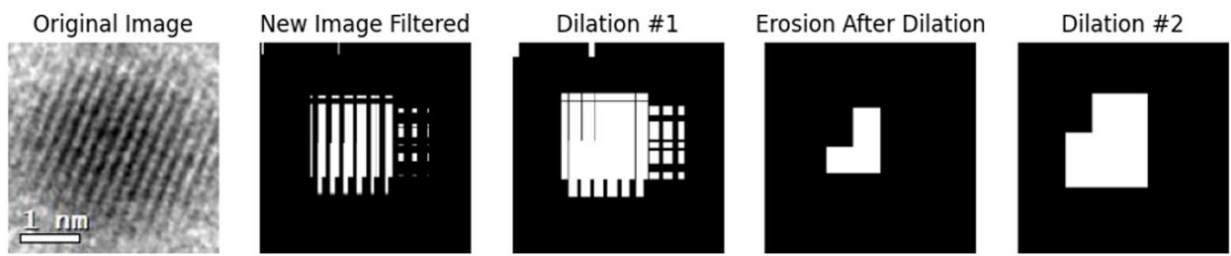

*Figure 5. In the second image from the left, statistical operations are executed along both rows and columns of the thresholded image. The initial steps involve dilation and erosion to eliminate undesired components, while the subsequent dilation focuses on enlarging the components that remain after the initial processing steps.*

A function of Opencv to identify connected white components within a binary image is utilized. The bounding box generated around these components is then scaled by a user-defined factor for further enlargement. Alternatively, an increased kernel size value for the final dilation operation can achieve a similar enlargement. The region of the gray image that corresponds to the location of the bounding box is subjected to skeletonization using the previously mentioned function. Subsequently, black-to-black distances are computed horizontally for each row, representing the atomic arrays and the local displacements, which can be used for calculating strains at the unit-cell length-scale. The images resulting from these steps are presented in **Fig. 6.**

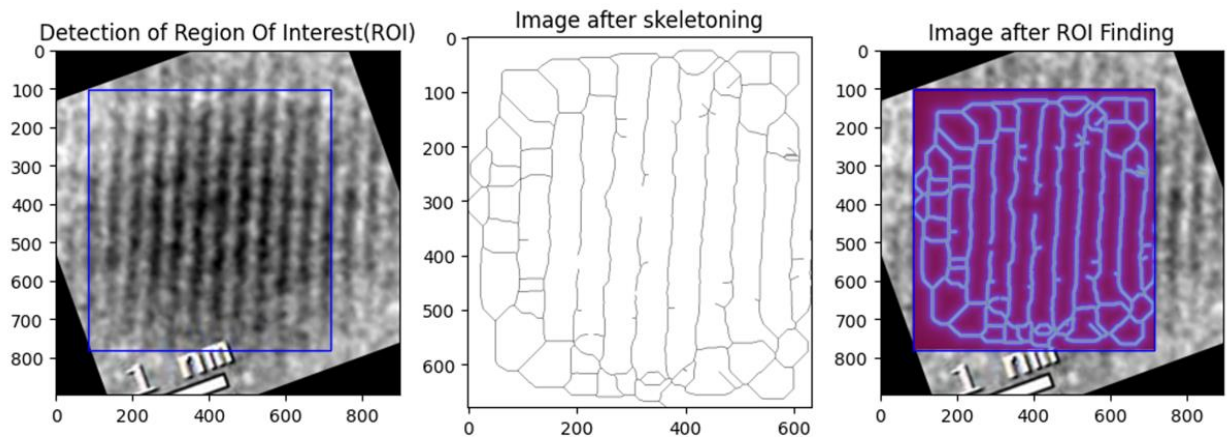

*Figure 6. The identified region of interest, the outcome after skeletonization, and the recolored version overlaid onto the entire image. Users have the flexibility to adjust the magnification factor, allowing for personalized enlargement of the identified region.*

Based on the extracted distances, we devised a customized color mapping scheme to assign colors to specific sequences of black-to-black pixels. The reference point for this mapping was established using a user-provided reference distance value. The actual distances of these sequences were then determined by correlating pixel counts with the pre-established count that corresponds to the length of the ruler.

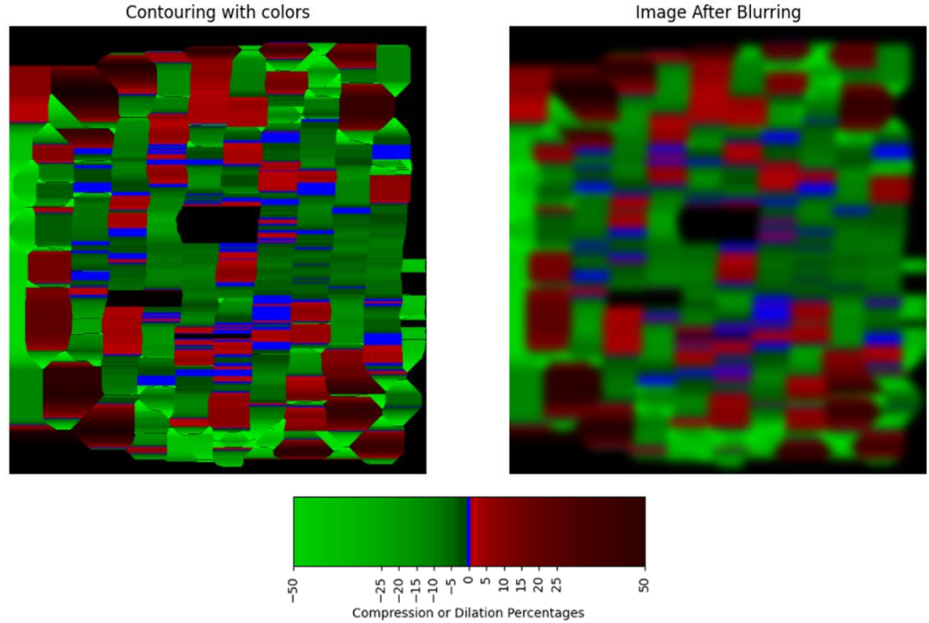

*Figure 7.* *The extracted image within the region of interest is color-coded using a custom colormap determined by the user-supplied reference distance value. The blurring operation is applied to the image to create a smoother transition at the edges of the color sections.*

The user has the option to generate an Excel file containing data on bin-count corresponding to the extracted compression or dilation percentages. Following this, the colored and blurred image is rotated again using the negation of the user-entered angle. This process is executed to achieve the final-colored image superimposed onto the original image.

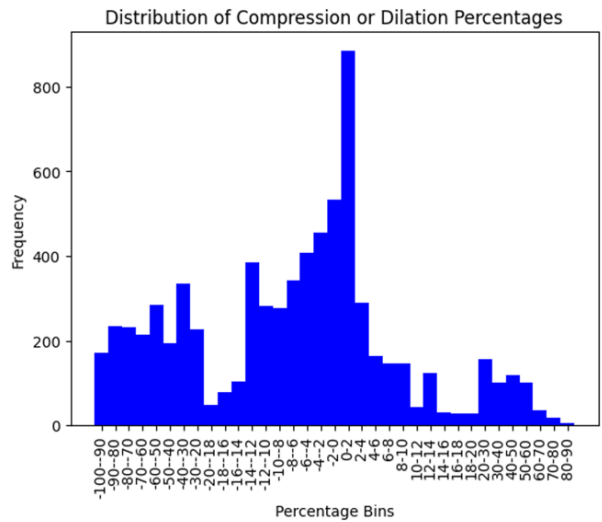

*Figure 8.* *Histogram depicting the statistical distribution of local lattice compression and expansion percentages.*

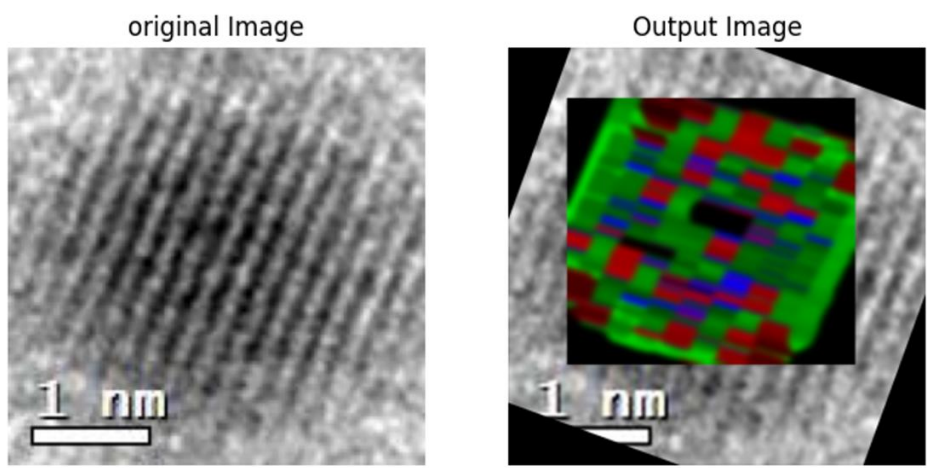

*Figure 9. The initial original image and the resulting output image after the application of all image processing steps. Users have the ability to interactively adjust flexible parameters to enhance the quality of the output image.*

As shown in **Fig. 7**, **8,** and **9**, we observe significant lattice expansion and compression locally that exceed 20% of the bulk lattice parameters, which raises the question of what changes in materials properties should be expected under such extreme strains and highlights the need for additional research on strain induced materials design at the nanoscale and beyond. Previous research has shown the tremendous importance of such atomic scale strains in material degradation [15], as well as startling changes in magnetic and optical characteristics [16]. As a result, PyNanospacing can assist materials engineers in better comprehending the cause and effect chain of reasoning from the standpoint of strain-induced phenomena.

**Conclusion**

We developed a software application in Python designed for accurate calculations and analyses of local lattice strains using TEM images. PyNanospacing application allows users to interact with important image processing parameters to create contour maps that show lattice stresses and variations in interplanar spacing in a variety of materials, including nanoparticles, 2D materials, and solid solutions. The software simplifies the visualization of lattice expansions and compressions by using statistical analysis and contour data. We believe that this application will positively impact the field of strain-engineered materials, securing a brighter future in strain engineering, ahead.

**Acknowledgements**

We gratefully acknowledge the European Research Council (ERC) funding for the Quantum Super-Exchange Energy Storage Platform (QUEEN) project, ERC-2021-STG 101043219.

**CRediT authorship contribution statement**

M.A.S. carried out the coding, developed methodology, prepared the visualizations and drafted the original manuscript. M.M. conceptualized the study, and helped in method development and writing of the manuscript. M.S. took part in conceptualization of the study and writing of the manuscript. S.T. supervised and acquired the necessary materials for this study. M.Ü. supervised and assisted in the conceptualization of this research. O.E. supervised, conceptualized, assisted in writing and funded this study.